\documentclass{vldb}

\usepackage[pdftex]{hyperref}
\hypersetup{
pdfauthor={Arun Kejariwal}, 
pdfcreator={pdflatex},
pdfproducer={pdflatex}, 
pdfborder={0 0 0},
pdfpagemode=UseOutlines, 
colorlinks=true,
anchorcolor = JungleGreen,
filecolor = cyan,
menucolor=blue,
pagecolor = magneta,
citecolor = Magenta, 
linkcolor=red, 
urlcolor=blue,
bookmarks=true,
backref=true,
bookmarksopen,
bookmarksnumbered,
hyperindex=true
}

\usepackage[dvipsnames,usenames]{color}
\definecolor{webgreen}{rgb}{0, 0.5, 0} 
\definecolor{webblue}{rgb}{0, 0, 0.5} 
\definecolor{webred}{rgb}{0.5, 0, 0} 

\usepackage{pifont}
\renewenvironment{dinglist}[2][Blue]
{\begin{list}{\textcolor{#1}{\ding{#2}}}{}}{\end{list}}
\DeclareFontFamily{OT1}{pzc}{}%
\DeclareFontShape{OT1}{pzc}{m}{it}{<-> s * [1.200] pzcmi7t}{}%
\DeclareMathAlphabet{\mathpzc}{OT1}{pzc}{m}{it}

\usepackage{eucal}
\usepackage{mathrsfs}

\usepackage{subcaption}
\usepackage[pdftex]{floatflt}
\usepackage{graphicx}
\usepackage{multirow}
\usepackage{latexsym}

\usepackage{amssymb}
\usepackage{amsmath}
\usepackage{shadow}
\usepackage{amsbsy}
\usepackage{amsfonts}
\usepackage{enumerate}
\usepackage{moreverb}
\usepackage{array}

\usepackage{tikz}

\usepackage{fancyhdr}
\usepackage{colortbl}
\usepackage{xcolor}

\newlength{\additionalvspace}%
\newlength{\colwidth}%
\newlength{\colwidthA}%
\newlength{\colwidthB}%
\newlength{\colwidthBb}%
\newlength{\colwidthC}%
\def\figref#1{Figure~\ref{#1}}
\def\secref#1{Section~\ref{#1}}

\def\tabref#1{Table~\ref{#1}}

\usepackage{graphicx}
\usepackage{tabulary}
\usepackage{hhline}
\usepackage{balance}  

\graphicspath{{../figs/}}

\newcommand*{\refname}{Bibliography}
\begin{document}

\title{Real Time Analytics: \\ Algorithms and Systems}

\author{
\alignauthor
Arun Kejariwal,\ \ Sanjeev Kulkarni,\ \ Karthik Ramasamy \\
       \affaddr{Twitter Inc.}\\
       \email{\{akejariwal, skulkarni, kramasamy\}@twitter.com}
}


\maketitle

\begin{abstract}
{\bf \large V}elocity is one of the {\bf 4 {\large V}s} commonly used to 
characterize Big Data \cite{4VsBigData}. In this regard, Forrester remarked 
the following in Q3 2014 \cite{Gualtieri14}:
{\em 
  ``The high velocity, white-water flow of data from innumerable real-time 
  data sources such as market data, Internet of Things, mobile, sensors, 
  clickstream, and even transactions remain largely unnavigated by most 
  firms. The opportunity to leverage streaming analytics has never been 
  greater."
} 
Example use cases of streaming analytics include, but not limited to:
(a) visualization of business metrics in real-time
(b) facilitating highly personalized experiences 
(c) expediting response during emergencies.
Streaming analytics is extensively used in a wide variety of domains such
as healthcare, e-commerce, financial services, telecommunications, energy 
and utilities, manufacturing, government and transportation.

In this tutorial, we shall present an in-depth overview of streaming 
analytics -- applications, algorithms and platforms -- landscape. We 
shall walk through how the field has evolved over the last decade and 
then discuss the current challenges -- the impact of the other three
{\bf V}s, viz., {\bf V}olume, {\bf V}ariety and {\bf V}eracity, on 
Big Data streaming analytics. 
The tutorial is intended for both researchers and practitioners in the 
industry. We shall also present state-of-the-affairs of streaming analytics 
at Twitter.  
\end{abstract}

\vspace*{-2mm}
\section{Introduction} \label{sec:intro}

\noindent
Big Data is characterized by the increasing volume (of the order of zetabytes), 
and the velocity of data generation \cite{NSFBigData,Manyika11}. It is projected
that the market size of Big Data will climb up from the current market size of 
\$5.1 billion to \$53.7 billion by 2017 \cite{BigDataMarket}. In recent years, 
Big Data analytics has been transitioning from being predominantly offline (or 
batch) to primarily online (or streaming). The trend is expected to become 
mainstream owing to the various facets, exemplified below, of the emerging 
data-driven society \cite{Pentland13}.

\newcolumntype{s}{>{\small}L}
\begin{table*}[!t]
\hspace*{-10mm}
\begin{tabulary}{1.07\textwidth}{s|s|s} 
{\bf Problem} & \multicolumn{1}{c|}{\bf Description} 
              & \multicolumn{1}{c}{\bf Application} \\ \hline\hline

Sampling \cite{Babcock02,Zhang05,Cormode05a,Aggarwal06,Tao07,Gandhi07,Gemulla08,Braverman09,Cormode10,Cormode12} &  
Obtain a representative set of the stream & 
A/B Testing \\ \hline

Filtering \cite{Bloom70,Cohen03,Pagh05,Donnet06,Bonomi06,Hao07,Kirsch08,Rothenberg10,Putze10,Moraru12,Dautrich13,Zhang13,Rottenstreich14,Fan14} & 
\multirow{2}{*}{Extract elements which meet a certain criterion} & 
Set membership \\ \hline

\multirow{2}{*}{Correlation \cite{Wang03,Sayal04,Pan12,Xie13,Guo14}} &  
Find data subsets (subgraphs) in (graph) data stream which are highly correlated
to a given data set & 
\multirow{2}{*}{Fraud detection} \\ \hline

Estimating Cardinality \cite{Flajolet83,Yossef02,Durand03,Giroire05,Flajolet07,Ganguly07,Chabchoub10,Kane10,Heule13,Chen11,Ting14} & 
\multirow{2}{*}{Estimate the number of distinct elements} & 
\multirow{2}{*}{Site audience analysis} \\ \hline 

Estimating Quantiles \cite{Greenwald01,Arasu04,Zhang07,Guha09,Hung10,Ma13,Shrivastava04} &
Estimate quantiles of a data stream with small amount of memory &
\multirow{1}{*}{Network analysis} \\ \hline

Estimating Moments \cite{Alon96,Coppersmith04,Indyk05,Bhuvanagiri06,Guha06} & 
Estimating distribution of frequencies of different elements & 
Databases \\ \hline

Finding Frequent Elements \cite{Manku02,Demaine02,Karp03,Cormode03,Jin03,Charikar04,Cormode05,Metwally05,Cormode08,Tanbeer09,Tanbeer09a,Manerikar09,Feigenblat10,Hung10a,Homem10,Yang11,Sahpaski12,Calders14,Pripuzic15} & 
\multirow{3}{*}{Identify items in a multiset with frequency more than a threshold $\theta$} & 
\multirow{3}{*}{Trending Hashtags} \\ \hline

\multirow{3}{*}{Counting Inversions \cite{Ajtai02}} &
\multirow{3}{*}{Estimate number of inversions} & 
Measure sortedness of data  \\ \hline

\multirow{2}{*}{Finding Subsequences \cite{Nowell05,Sun07,Gal10,Toyoda13}} &
Find Longest Increasing Subsequences (LIS), Longest Common Subsequence (LCS),
subsequences similar to a given query sequence &
\multirow{2}{*}{Traffic analysis} \\ \hline

\multirow{2}{*}{Path Analysis \cite{Eppstein97}} &
Determine whether there exists a path of length $\leq \ell$ between two nodes in a dynamic graph &
Web graph analysis \\ \hline

Anomaly Detection \cite{Papadimitriou05,Subramaniam06,Su07,Khan09,Dasu09,Teixeira10,Beigi11,Tan11,Assent12} & 
\multirow{2}{*}{Detect anomalies in a data stream} &
\multirow{2}{*}{Sensor networks} \\ \hline

Temporal Pattern Analysis \cite{Chen07,Yu12,Alavi14} & 
Detect patterns in a data stream & 
\multirow{1}{*}{Traffic analysis} \\ \hline 

\multirow{2}{*}{Data Prediction \cite{Kalman60,Wan00,Halatchev05,Wang05,Rodrigues06,Vijayakumar07}} & %
\multirow{2}{*}{Predict missing values in a data stream} & 
Sensor data analysis  \\ \hline

Clustering \cite{Guha00,Callaghan02,Hore07} &  
Cluster a data stream & 
\multirow{1}{*}{Medical imaging} \\ \hline

\multirow{3}{*}{Graph analysis \cite{Feigenbaum05,Halldorsson10,Ahn12,Kapralov14,McGregor14,Chitnis15,Esfandiari15}} &  
 Extract unweighted and weighted matching, vertex cover, independent sets, spanners, subgraphs (sparsification) and random walks, computing min-cut & 
\multirow{3}{*}{Web graph analysis} \\ \hline 

\multirow{2}{*}{Basic Counting \cite{Datar02}} & 
Estimate $\hat{m}$ of the number $m$ of 1-bits in the sliding window 
(of size n) such that $|\hat{m} - m| \leq \epsilon m$ & 
\multirow{2}{*}{Popularity Analysis} \\ \hline 

\multirow{2}{*}{Significant One Counting \cite{Lee06}} & 
Estimate $\hat{m}$ of the number $m$ of 1-bits in the sliding window (of 
size n) such that if $m \geq \theta n$, then $|\hat{m} - m| \leq \epsilon m$ & 
\multirow{2}{*}{Traffic accounting \cite{Estan02}} \\ \hline\hline 

\end{tabulary}
\vspace*{-3mm}
\caption{Streaming algorithms and their applications}
\label{tab:algos}
\vspace*{-5mm}
\end{table*}

\vspace*{-2mm}
\begin{dinglist}{122} 
\item Social media: Over 500M tweets are created everyday. A key challenge in
      this regard is how to surface the most personalized content in real time.
      \vspace*{-2mm}
\item Internet of Things (IoT): By 2020, the number of connected devices is 
      expected to grow by 50\% to 30 billion \cite{IoT1}. Data from embedded 
      systems -– the sensors and systems that monitor the physical universe -– 
      is expected to rise to 10\% (from the current 2\%) of the digital universe
      by 2020.
      \vspace*{-2mm}
\item Health Care: Increasingly Big Data is being leveraged in health care to, 
      for example, improve both quality and efficiency in health care areas 
      such as readmissions, adverse events, treatment optimization, and early
      identification of worsening health states or highest-need populations
      \cite{HealthCare}. The volume of healthcare data is expected to swell to 
      2,314 exabytes by 2020, from 153 exabytes in 2013 \cite{HealthCare1}.
      \vspace*{-2mm}
\item Machine data: With cloud computing becoming ubiquitous, machine generated
      data is expected to grow to 40\% of the digital universe by 2020 
      \cite{EMCReport}.
      \vspace*{-2mm}
\item Connected vehicles: New telematics systems and the installation of ever
      greater numbers of computer chips, applications, electronic components and
      many other components provide data on vehicle usage, wear and tear,
      or defects \cite{Auto}. The volume of data transferred per vehicle per
      month is expected to grow from around 4 MB to 5 GB. Further, by 2016 as
      many as 80\% of all vehicles sold worldwide are expected to be ``connected". 
\end{dinglist}
\vspace*{-2mm}

\begin{figure}[!t]
\centering 
\includegraphics[width=0.90\linewidth]{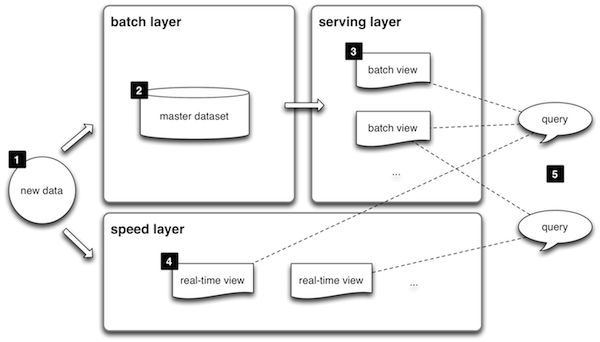}
\vspace*{-3mm}
\caption{Overview of Lambda Architecture (source: \protect\cite{Lambda})}
\vspace*{-5mm}
\label{fig:lambda}
\end{figure}

\noindent
Over the years, several streaming platforms have been developed. Examples include, 
S4 \cite{S4}, Samza \cite{Samza}, Sonora \cite{Yang12}, Millwheel \cite{Akidau13}, 
Photon \cite{Ananthanarayanan13}, Storm \cite{Storm}, Flink \cite{Flink}, Spark 
\cite{Spark}, Pulsar \cite{Pulsar} and Heron \cite{Kulkarni15}. Some of these 
platforms have been open sourced. The evolution of the streaming platforms is 
discussed in detail in \secref{plat}. 

In order to be satisfy both, batch and streaming analytics, Lambda Architecture
(LA) has been proposed as a robust, distributed platform to serve a variety of 
workloads, including low-latency high-reliability queries \cite{Lambda} (refer 
to \figref{fig:lambda}). The various stages of LA are explained below:

\begin{enumerate}
\item Input {\bf data} is dispatched to both the batch layer and the speed layer 
      for processing.
      \vspace*{-2mm}
\item The {\bf batch layer} manages the master dataset (an immutable, append-only 
      set of raw data) and pre-computes the batch views.
      \vspace*{-2mm}
\item The {\bf serving layer} indexes the batch views so that they can be queried
      in a low-latency, ad-hoc way.
      \vspace*{-2mm}
\item The {\bf speed layer} handles recent data only to compensate for the high 
      latency of updates to the serving layer.
      \vspace*{-2mm}
\item Incoming {\bf queries} are answered by merging results from batch views and 
      real-time views.
\end{enumerate}
\vspace*{-2mm}




\noindent 
Several platforms have been built based on the Lambda Architecture. Examples
include Summingbird \cite{Summingbird} and Lambdoop \cite{Lambdoop}. 
Commercial platforms such as TellApart \cite{TellApart} are also based on the 
Lambda Architecture.

%

The rest of the proposal is organized as follows: 
\secref{over} overviews the various problems addressed previously in the context
of streaming analytics and their real-world applications. 
\secref{plat} walks through the evolution of streaming platforms over the last
decade. 
Finally, we conclude in \secref{sec:conclusions}.

\vspace*{-2mm}
\section{Streaming Algorithms} \label{over}

\noindent
Elements of a data stream need to be processed in real time, else one may lose 
the opportunity to process them at all. Thus, it is critical that the data
footprint of the algorithm fits in the main memory. Also, in light of the 
real-time constraint, it may be preferable to compute an approximate solution 
than an exact solution. Research in approximation algorithms for problems 
defined over data streams has led to some general techniques for data 
reduction and synopsis construction, including:

\vspace*{-1mm}
\begin{dinglist}{122} 
\item {\bf \em Sampling}: Techniques such as reservoir sampling 
      \cite{Vitter85}, weighted sampling \cite{Chaudhuri99} have been 
      proposed to capture the essential characteristics of a data stream.
      \vspace*{-2mm}
\item {\bf \em Sliding windows}: Use of sliding windows prevents stale data 
      from influencing analysis and statistics and also serve as a tool for 
      approximation, given bounded memory.
      The following problems for sliding windows are being actively researched: 
      clustering, 
      maintaining statistics 
      like variance, and computing correlated aggregates.
      \vspace*{-2mm}
\item {\bf \em Clustering}: Algorithms for problems such as the $k$-median 
      problem -- wherein the objective is to choose $k$ representative points, 
      such that the sum of the errors over the $n$ data points is minimized -- 
      have been proposed based on clustering \cite{Guha00,Guha09a,Silva13,
      Aggarwal13}.
      \vspace*{-2mm}
\item {\bf \em Sketches}: Randomized sketching, introduced by Alon et al.\ 
      \cite{Alon96}, summarizes a data stream using a small amount of memory. 
      The sketch if used to estimate the answer to certain queries (typically,
      ``distance" queries) over a data set.
      \vspace*{-2mm}
\item {\bf \em Histograms}: V-Optimal histogram approximates the distribution 
      of a set of values $v_1, \ldots v_n$ by a piecewise-constant function
      $\hat{v}(i)$, so as to minimize the sum of squared error. Equi-width
      histograms partition the domain into buckets such that the number of 
      $v_i$ values falling into each bucket is uniform across all buckets.
      End-biased histograms maintain exact counts of items that occur with 
      frequency above a threshold, and approximate the other counts by a 
      uniform distribution.
      \vspace*{-5mm}
\item {\bf \em Wavelets}: Wavelets coefficients are projections of the given 
      signal (set of data values) onto an orthogonal set of basis vectors. The 
      coefficients have the desirable property that the signal reconstructed 
      from the top few wavelet coefficients best approximates the original 
      signal in terms of the $L_2$ norm \cite{Gilbert02}. The choice of basis 
      vectors determines the type of wavelets. 
\end{dinglist}
\vspace*{-2mm}

\begin{table*}
\centering
\begin{tabulary}{\textwidth}{p{2cm}|s} \hline
{\bf Platform} & \multicolumn{1}{c}{\bf Description} \\ \hline\hline

S4 \cite{S4} & 
Real-time analytics with a key-value based programming model and support for
scheduling/message passing and fault tolerance \\ \hline

Storm \cite{Storm}  & 
The most popular and widely adopted real-time analytics platform developed at
Twitter \\ \hline

Millwheel \cite{Akidau13} & 
Google's proprietary realtime analytics framework thats provides exact once
semantics
\\ \hline

Samza \cite{Samza} & 
Framework for topology-less real-time analytics that emphasizes sharing between
groups \\ \hline

Akka \cite{Akka} &
Toolkit for writing distributed, concurrent and fault tolerant applications \\ \hline

Spark \cite{Spark}  & 
Does both offline and online analysis using the same code and same system \\ \hline

Flink \cite{Flink} & 
Fuses offline and online analysis using traditional RDBMS techniques \\ \hline

Pulsar \cite{Pulsar} & 
Does real-time analytics using SQL \\ \hline

Heron \cite{Kulkarni15} & 
Storm re-imagined with emphasis on higher scalability and better debuggability \\ \hline\hline

\end{tabulary}
\vspace*{-2mm}
\caption{Open source streaming platforms}
\label{tab:plats}
\vspace*{-5mm}
\end{table*}

\noindent 
Further, to be able to support Web scale and high velocity data, the algorithms
should intrinsically distribute computation across multiple nodes and, if 
required, across data centers. In other words, the algorithms should be able 
to scale out. 

In light of the dynamic nature of streaming data, a field of incremental machine
learning has emerged to cater to Big Data streaming analytics. The techniques 
being developed are designed to work with incomplete data, to identify hidden 
variables to help steer future data collection and to quantify the change between 
one or more states of the model.

Common streaming operators include, but not limited to: filtering, time windows,
aggregation/correlation, temporal patterns, location/motion, enrichment, query
and action interfaces.
\tabref{tab:algos} lists some of the most common problems addressed in prior
research in the domain of streaming analytics and their example applications 
in the real world. Examples of use of streaming analytics include, but not
limited to, 
(a) sequence mining \cite{Raissi08,Li09,Koper11} for, say, credit card fraud 
    detection, motion capture sequences and chlorine levels in drinking water
(b) discovering human activity, which often exhibit discontinuity (interruption) 
    or varying frequencies, from sensor streams \cite{Rashidi10} 
(c) determing top-K traversal sequences in streaming clicks 
(d) finding closed structures in music melody streams \cite{Li04}. 
In December 2015, yahoo! open source a library called {\em DataSketches} for
approximate analysis of Big Data \cite{Rhodes15}.

We shall walk through some of the problems and the recent algorithms in the 
tutorial. Further, we shall throw light on the scalability of the existing 
approaches at Web scale. 

\vspace*{-2mm}
\section{Streaming Platforms} \label{plat}

\noindent
In late 1990s and early 2000s, main memory DataBase Management Systems (DBMSs) and 
rule engines\footnote{A rule engine typically accepts condition/action pairs, usually
expressed using ``if-then'' notation. As streaming data enters the system, it is
immediately matched against the existing rules.  When the condition of a rule is
matched, the rule is said to ``fire". The corresponding action(s) taken may then
produce alerts/outputs to external applications or may simply modify the state
of internal variables, which may in turn lead to further rule firings.}, were
re-purposed and remarketed to cater to stream processing. However, these systems 
did not scale with high volume data streams (models and issues in data stream 
systems are discussed in detail in \cite{Babcock02a}).
Later on, Stream Processing Engines (SPEs) such as Aurora \cite{Carney02}, STREAM 
\cite{Arasu03}, TelegraphCQ \cite{Chandrasekaran03} and Borealis \cite{Abadi05} were 
proposed. Even these systems did not scale with the increasing velocity and volume 
of the data streams characteristic of modern systems. To this end, several streaming 
platforms have been developed in the industry. 
\tabref{tab:plats} summarizes the various streaming platforms developed over the 
years. 

Some of the common requirements of streaming systems are itemized below:

\vspace*{-2mm}
\begin{dinglist}{122} 
\item Provide resiliency against stream ``imperfections", including missing and 
      out-of-order data, which are commonly present in data streams in production.
      \vspace*{-2mm}
\item Must guarantee predictable and repeatable outcomes.
      \vspace*{-3mm}
\item Ensure that the applications are up and available, and the integrity of the 
      data is maintained at all times despite failures (which can happen due to, 
      for example, node failures, network failures, software bugs and resource 
      limitations \cite{Hwang05}). 
      \vspace*{-2mm}
\item Distribute processing across multiple processors and machines to achieve 
      incremental scalability.
      \vspace*{-2mm}
\item Should be easy to integrate with a batch processing data pipeline (ala the 
      Lambda architecture described in \secref{sec:intro}). This is key for a  
      wide variety of applications, such as online fraud detection, electronic 
      trading based on historical patterns.
\end{dinglist}
\vspace*{-2mm}

\noindent
In the rest of this section, we briefly overview the platforms listed in 
\tabref{tab:plats}. In addition, we also overview low-latency platforms built on 
top of Hadoop. In the tutorial, we shall walk the audience through the different
design choices of the various platforms and the challenges which still remain.

S4 \cite{S4} is one of earliest distributed streaming system developed by Yahoo! It is near 
real-time, scalable and event-driven platform that allows easy implementation of
applications for processing unbounded streams of data. At a high level, it allows 
for easy assembly of small applications into larger ones, flexible and easy 
deploy, provides fault tolerance for high availability, checkpointing and a
recovery mechanism for minimizing state loss.  The platform handles communication, 
scheduling and distribution. S4 streaming applications are modeled as a graph, 
with vertices representing computation (called processing elements) and the edges 
representing streams of data. The applications are deployed on S4 clusters that
run several distributed containers called S4 nodes. Processing elements communicate 
asynchronously by sending events on streams. These events are routed to the 
appropriate nodes according to their key.

Apache Storm \cite{Storm} is the next generation system that is widely popular and open sourced 
by Twitter.  Storm applications, referred to as topologies, is a DAG where the 
vertices can either represent a data source (spouts) and a computation (bolts). 
These topologies are run on a Storm cluster.  Storm provides guarantees about data 
processing with support for {\em at least once} and {\em almost once} semantics. It 
is horizontally 
scalable thereby allowing the cluster to expand and supports robust fault tolerance 
for process and machine failures. Storm data model allows users to express their 
analytics concisely. A Storm cluster consists of Nimbus that acts as a master
node and is responsible for scheduling and distribution of topologies. Other nodes
in the cluster, called Slave Nodes, run Storm Supervisor that spawns workers 
which actually run the user logic code.

MillWheel \cite{Akidau13} is a key-value based streaming system developed at Google.
A MillWheel application is a directed graph where each node is a computational
unit and the vertices are the messages passed between them. MillWheel distributes 
the computational nodes across the cluster and repairs them in case units/machines 
go down. Mill wheel also provides {\em exactly once} semantics by checkpointing 
state every time. To checkpoint reliably, MillWheel uses BigTable \cite{Chang06}. 
MillWheel's programming model provides a notion of logical time, making it simple 
to write time-based aggregations. MillWheel is closed-source.

Apache Samza \cite{Samza} is a realtime, asynchronous computational framework for stream
processing developed at LinkedIn. Unlike Storm or MillWheel, where you stitch
together a bunch of computations in a topology, a Samza application is single
computational task scaled across several partitions. Each Samza application
reads one or more input streams and can output zero or more output streams. One
can then stitch together several such applications to form a Storm like topology
doing a given higher level function. Samza uses Kafka \cite{Kafka} to manage the
input and output streams. One side-effect of using Kafka for stream management 
is that Samza inherits all the persistence and fault-tolerance of Kafka. As all 
streams exist on Kafka, one does not need external systems/brokers for 
inter-application communication. However this comes at the cost of increased 
latency as even the intermediate stages have to be persisted to disk.

Akka \cite{Akka} is a toolkit for building distributed, concurrent and fault-tolerant
applications. One can use Akka to build general data processing applications -- 
batch or streaming. An Akka application consists of a set of Akka Actors and 
messages passed between those Actors. An Akka Actor is very
similar to a Storm Bolt, except that it is very lightweight. Thus, it is
commonplace to see millions of Akka Actors in a single Akka application. Actors
process messages asynchronously and each actor instance is guaranteed to be run
using at most one thread at a time, making concurrency much easier. Akka
provides out-of-the-box primitives to distribute actors across the cluster,
do load balancing of messages and repair lost actors. A unique feature of Akka
is that actors can reply to incoming messages thereby giving it a
request-response capability thats usually not present in systems.

Apache Spark \cite{Spark} is an effort that came out of AMPLabs Berkeley to replace Hadoop's
two stage disk-based MapReduce paradigm. Spark provides in-memory primitives
which allow intermediate data to be kept in memory. Spark distributes Resilient
Distributed Datasets (RDDs) throughout the cluster and can even store them to 
disk for persistence. For a class of iterative machine learning algorithms, this 
in-memory approach provides $100\times$ more throughput than traditional MapReduce 
based implementations. As a result, the community has built a large set of ML and
Graph processing libraries on top of Spark.
APIs are provided in Java/Python/Scala languages. Spark also provides streaming
primitives so that streaming applications can run in the same cluster as batch
applications. This consolidation of infrastructure for running disparate classes
of applications drives down the opex cost significantly. Spark Streaming provides 
a high-level abstraction called discretized stream or DStream, which represents 
a continuous stream of data. DStreams can be created either from input data 
streams from sources such as Kafka, Flume \cite{Flume}, Twitter, ZeroMQ
\cite{ZeroMQ}, Kinesis \cite{Kinesis} or TCP sockets 
or by applying high-level operations on other DStreams (internally, a DStream is 
represented as a sequence of RDDs). The data can be processed using complex
algorithms expressed with high-level functions like map, reduce, join and
window. Finally, processed data can be pushed out to filesystems, databases, and
live dashboards. Spark streaming supports stateful {\em exactly once} semantics 
out-of-the-box.

Apache Flink \cite{Flink} takes a different approach to achieve the same goal as Spark. Flink
borrows concepts from the traditional RDBMS world like byte-buffer based data 
serialization and binary representation of data (instead of Java/Scala object 
representation). Flink has a cost-based optimizer, akin to  relational platforms 
that selects execution strategies and avoids expensive partitioning and sorting 
steps. Moreover, Flink features a special kind of iterations called delta-iterations 
that can significantly reduce the amount of computations as iterations go on. Like 
Spark, Flink also unifies stream and batch processing.

Pulsar \cite{Pulsar} is a realtime analytics engine open sourced by eBay. A unique feature of
Pulsar is its SQL interface. Thus, instead of writing code in, say, Java, one 
can just write SQL queries to run on a Pulsar cluster. This eases the use of the 
analytics pipeline by non-technical business folks who tend to know SQL pretty 
well. Pulsar transforms each query into a directed acyclic graph (DAG) of 
processing nodes and distributes them across the cluster. Pulsar achieves low 
latencies by keeping all intermediate data in memory. However if the downstream 
components are either down or not able to consume fast enough, it stores the 
messages into Kafka for later replay. Another neat feature of Pulsar is the 
ability to dynamically resize queries on the fly while the query is still running. 
In this way, one can add/remove machines into a Pulsar cluster without affecting 
any running queries.

After years of experience with Storm, as the scale of data being processed in real-time 
increased, several issues such as debugability, manageability, scalability and
performance became apparent. Most of these issues were a result of the underlying
architectural issues such as multiplexing of disparate tasks running user logic code 
in a single worker process.  As a consequence, a worker has a complex set of queues 
through which the data passes making the performance worse. Heron \cite{Kulkarni15} 
addresses these issues by running each task in a process of its own thereby making 
it easy to debug, tune and improved performance.

While Apache Hive \cite{Hive} opened up HDFS to SQL, its architecture, centered
around MapReduce \cite{Dean04}, made it unsuitable for interactive querying. 
Quite a few efforts have been initiated by different companies to solve this
problem. The most prominent ones are Drill \cite{Drill} from MapR, Presto
\cite{Presto} from Facebook, Impala \cite{Impala} from Cloudera and Tez
\cite{Tez} from Hortonworks. While differing in details, they all generally
have the same architecture. All of them prefer to be co-located with the HDFS
for best performance. An incoming SQL query is parsed and a physical plan is 
generated which is then optimized. A query co-ordinator sends pieces of the query 
to all relevant data nodes where servers execute that part of query, reading from
the local data node if needed. This is done to minimize network traffic. The
query co-ordinator then merges all the results and returns the combined result
back to the user. The systems differ in the flavor of supported SQL (while Presto
and Drill support ANSI SQL, Impala supports HiveQL), language of implementation
(Impala is written in C++, while others are all Java) and levels of
maturity/adoption. 

In addition to the platforms discussed above, several commercial stream processing 
products are available on the market \cite{Infosphere,Vibe,ESP,Apama,Streambase,
Blaze,Vitria}. 
In \cite{Gualtieri14}, Gualtieri and Curran reviewed some of the widely used 
and emerging commercial streaming analytics platforms. Numenta has developed 
a tool, called {\em Grok} \cite{Grok}, for anomaly detection in data streams.  

Lastly, we shall walk the audience through the various use cases of Heron for
streaming analytics -- such as, but not limited to, real-time targeting, content 
discovery, online machine learning -- at Twitter. 

\section{Conclusions} \label{sec:conclusions}

\noindent
In the proposed tutorial, we shall present an in-depth overview of
streaming analytics -- applications, algorithms and platforms -- 
landscape. We shall walk through how the field has evolved over the 
last decade and then discuss the current challenges. 

{%
\fontsize{5}{0.19cm}%
\selectfont%
\bibliographystyle{abbrv}

\begin{thebibliography}{100}

\bibitem{Akka}
{Akka}.
\newblock \url{http://akka.io/}.

\bibitem{Kinesis}
{Amazon Kinesis}.
\newblock \url{http://aws.amazon.com/kinesis/}.

\bibitem{Drill}
{Apache Drill}.
\newblock \url{http://drill.apache.org/}.

\bibitem{Flink}
{Apache Flink}.
\newblock \url{https://flink.apache.org/}.

\bibitem{Flume}
{Apache Flume}.
\newblock \url{http://flume.apache.org/}.

\bibitem{Hive}
{Apache Hive}.
\newblock \url{https://hive.apache.org/}.

\bibitem{Kafka}
{Apache Kafka: A high-throughput distributed messaging system}.
\newblock \url{http://kafka.apache.org/}.

\bibitem{Samza}
{Apache Samza}.
\newblock \url{https://samza.apache.org/}.

\bibitem{Spark}
{Apache Spark}.
\newblock \url{https://spark.apache.org/}.

\bibitem{Tez}
{Apache Tez}.
\newblock \url{https://tez.apache.org/}.

\bibitem{Apama}
{Apama Streaming Analytics}.
\newblock
  \url{http://www.softwareag.com/corporate/products/apama_webmethods/analytics/overview/default.asp}.

\bibitem{BigDataMarket}
{Big Data Market Size and Vendor Revenues}.
\newblock
  \url{http://wikibon.org/wiki/v/Big_Data_Market_Size_and_Vendor_Revenues}.

\bibitem{Auto}
{Connected cars get big data rolling}.
\newblock \url{http://www.telekom.com/media/media-kits/179806}.

\bibitem{Grok}
{Grok}.
\newblock \url{http://numenta.com/grok/}.

\bibitem{Impala}
{Impala}.
\newblock \url{http://impala.io/}.

\bibitem{Vibe}
{Informatica Vibe Data Stream}.
\newblock
  \url{https://www.informatica.com/products/data-integration/real-time-integration/vibe-data-stream.html#fbid=Z8rhqt-b1nd}.

\bibitem{Infosphere}
{InfoSphere Streams: Capture and analyze data in motion}.
\newblock \url{http://www-03.ibm.com/software/products/en/infosphere-streams}.

\bibitem{Lambda}
{Lambda Architecture}.
\newblock \url{http://lambda-architecture.net/}.

\bibitem{Lambdoop}
{Lambdoop}.
\newblock \url{http://lambdoop.com/}.

\bibitem{EMCReport}
{New Digital Universe Study Reveals Big Data Gap: Less Than 1\% of World’s
  Data is Analyzed; Less Than 20\% is Protected}.
\newblock \url{http://www.emc.com/about/news/press/2012/20121211-01.htm}.

\bibitem{Presto}
{Presto}.
\newblock \url{https://prestodb.io/}.

\bibitem{ESP}
{SAP Event Stream Processor}.
\newblock
  \url{http://www.sap.com/pc/tech/database/software/sybase-complex-event-processing/index.html}.

\bibitem{Blaze}
{SQLstream Blaze}.
\newblock \url{http://www.sqlstream.com/blaze/}.

\bibitem{Summingbird}
{Summingbird}.
\newblock \url{https://github.com/twitter/summingbird}.

\bibitem{TellApart}
{TellApart}.
\newblock \url{http://www.tellapart.com}.

\bibitem{IoT1}
{The Digital Universe of Opportunities: Rich Data and the Increasing Value of
  the Internet of Things}.
\newblock
  \url{http://www.emc.com/leadership/digital-universe/2014iview/internet-of-things.htm}.

\bibitem{4VsBigData}
{The Four V's of Big Data}.
\newblock \url{http://www.ibmbigdatahub.com/infographic/four-vs-big-data}.

\bibitem{Streambase}
{TIBCO StreamBase}.
\newblock \url{http://www.streambase.com/}.

\bibitem{Vitria}
{Vitria OI For Streaming Big Data Analytics}.
\newblock
  \url{http://www.vitria.com/solutions/streaming-big-data-analytics/benefits/}.

\bibitem{ZeroMQ}
{ZeroMQ}.
\newblock \url{http://zeromq.org/}.

\bibitem{NSFBigData}
{Federal Government Big Data Rollout}.
\newblock
  \url{http://www.nsf.gov/news/news_videos.jsp?cntn_id=123607&media_id=72174&org=NSF},
  2012.

\bibitem{Abadi05}
D.~J. Abadi, Y.~Ahmad, M.~Balazinska, M.~Cherniack, J.~hyon Hwang, W.~Lindner,
  A.~S. Maskey, E.~Rasin, E.~Ryvkina, N.~Tatbul, Y.~Xing, and S.~Zdonik.
\newblock The design of the {Bo}realis stream processing engine.
\newblock In {\em Proceedings of the Conference on Innovative Data Systems
  Research}, pages 277--289, 2005.

\bibitem{Aggarwal06}
C.~C. Aggarwal.
\newblock On biased reservoir sampling in the presence of stream evolution.
\newblock In {\em Proceedings of the 32nd International Conference on Very
  Large Data Bases}, pages 607--618, Seoul, Korea, 2006.

\bibitem{Aggarwal13}
C.~C. Aggarwal.
\newblock A survey of stream clustering algorithms.
\newblock In C.~Aggarwal and C.~Reddy, editors, {\em Data Clustering:
  Algorithms and Applications, CRC Press}. 2013.

\bibitem{Ahn12}
K.~J. Ahn, S.~Guha, and A.~McGregor.
\newblock Graph sketches: Sparsification, spanners, and subgraphs.
\newblock In {\em Proceedings of the 31st Symposium on Principles of Database
  Systems}, pages 5--14, Scottsdale, AZ, 2012.

\bibitem{Ajtai02}
M.~Ajtai, T.~S. Jayram, R.~Kumar, and D.~Sivakumar.
\newblock Approximate counting of inversions in a data stream.
\newblock In {\em Proceedings of the Thiry-fourth Annual ACM Symposium on
  Theory of Computing}, pages 370--379, Montreal, Quebec, Canada, 2002.

\bibitem{Akidau13}
T.~Akidau, A.~Balikov, K.~Bekiro\u{g}lu, S.~Chernyak, J.~Haberman, R.~Lax,
  S.~McVeety, D.~Mills, P.~Nordstrom, and S.~Whittle.
\newblock Millwheel: Fault-tolerant stream processing at internet scale.
\newblock {\em Proceedings of the VLDB Endowment}, 6(11):1033--1044, Aug. 2013.

\bibitem{Alavi14}
F.~Alavi and S.~Hashemi.
\newblock Mining jumping emerging patterns by streaming feature selection.
\newblock In V.~N. Huynh, T.~Denoeux, D.~H. Tran, A.~C. Le, and S.~B. Pham,
  editors, {\em Knowledge and Systems Engineering}, volume 245 of {\em Advances
  in Intelligent Systems and Computing}, pages 337--349. 2014.

\bibitem{Alon96}
N.~Alon, Y.~Matias, and M.~Szegedy.
\newblock The space complexity of approximating the frequency moments.
\newblock In {\em Proceedings of the Twenty-eighth Annual ACM Symposium on
  Theory of Computing}, pages 20--29, Philadelphia, Pennsylvania, USA, 1996.

\bibitem{Ananthanarayanan13}
R.~Ananthanarayanan, V.~Basker, S.~Das, A.~Gupta, H.~Jiang, T.~Qiu,
  A.~Reznichenko, D.~Ryabkov, M.~Singh, and S.~Venkataraman.
\newblock Photon: Fault-tolerant and scalable joining of continuous data
  streams.
\newblock In {\em Proceedings of the 2013 ACM SIGMOD International Conference
  on Management of Data}, pages 577--588, 2013.

\bibitem{Arasu03}
A.~Arasu, B.~Babcock, S.~Babu, M.~Datar, K.~Ito, I.~Nishizawa, J.~Rosenstein,
  and J.~Widom.
\newblock {STREAM}: The stanford stream data manager (demonstration
  description).
\newblock In {\em Proceedings of the 2003 ACM SIGMOD International Conference
  on Management of Data}, pages 665--665, 2003.

\bibitem{Arasu04}
A.~Arasu and G.~S. Manku.
\newblock Approximate counts and quantiles over sliding windows.
\newblock In {\em Proceedings of the Twenty-third ACM Symposium on Principles
  of Database Systems}, pages 286--296, Paris, France, 2004.

\bibitem{Assent12}
I.~Assent, P.~Kranen, C.~Baldauf, and T.~Seidl.
\newblock Any{O}ut: Anytime outlier detection on streaming data.
\newblock In {\em Proceedings of the 17th International Conference on Database
  Systems for Advanced Applications - Volume Part I}, pages 228--242, 2012.

\bibitem{Babcock02a}
B.~Babcock, S.~Babu, M.~Datar, R.~Motwani, and J.~Widom.
\newblock Models and issues in data stream systems.
\newblock In {\em Proceedings of the Symposium on Principles of Database
  Systems}, pages 1--16, Madison, Wisconsin, 2002.

\bibitem{Babcock02}
B.~Babcock, M.~Datar, and R.~Motwani.
\newblock Sampling from a moving window over streaming data.
\newblock In {\em Proceedings of the Thirteenth Annual ACM-SIAM Symposium on
  Discrete Algorithms}, pages 633--634, San Francisco, California, 2002.

\bibitem{Yossef02}
Z.~Bar-Yossef, T.~Jayram, R.~Kumar, D.~Sivakumar, and L.~Trevisan.
\newblock Counting distinct elements in a data stream.
\newblock In D.~P. Rolim and S.~Vadhan, editors, {\em Randomization and
  Approximation Techniques in Computer Science}, volume 2483 of {\em Lecture
  Notes in Computer Science}, pages 1--10. 2002.

\bibitem{Beigi11}
M.~S. Beigi, S.-F. Chang, S.~Ebadollahi, and D.~C. Verma.
\newblock Anomaly detection in information streams without prior domain
  knowledge.
\newblock {\em IBM Journal of Research and Development}, 55(5):550--560, Sept.
  2011.

\bibitem{Bhuvanagiri06}
L.~Bhuvanagiri, S.~Ganguly, D.~Kesh, and C.~Saha.
\newblock Simpler algorithm for estimating frequency moments of data streams.
\newblock In {\em Proceedings of the Seventeenth Annual ACM-SIAM Symposium on
  Discrete Algorithm}, pages 708--713, Miami, Florida, 2006.

\bibitem{Bloom70}
B.~H. Bloom.
\newblock Space/{T}ime trade-offs in hash coding with allowable errors.
\newblock {\em Communications of the ACM}, 13(7):422--426, July 1970.

\bibitem{Bonomi06}
F.~Bonomi, M.~Mitzenmacher, R.~Panigrahy, S.~Singh, and G.~Varghese.
\newblock An improved construction for counting bloom filters.
\newblock In Y.~Azar and T.~Erlebach, editors, {\em Algorithms – ESA 2006},
  volume 4168 of {\em Lecture Notes in Computer Science}, pages 684--695. 2006.

\bibitem{Braverman09}
V.~Braverman, R.~Ostrovsky, and C.~Zaniolo.
\newblock Optimal sampling from sliding windows.
\newblock In {\em Proceedings of the Twenty-eighth ACM SIGMOD-SIGACT-SIGART
  Symposium on Principles of Database Systems}, pages 147--156, Providence,
  Rhode Island, USA, 2009.

\bibitem{Calders14}
T.~Calders, N.~Dexters, J.~J.~M. Gillis, and B.~Goethals.
\newblock Mining frequent itemsets in a stream.
\newblock {\em Information Systems}, 39:233--255, Jan. 2014.

\bibitem{Carney02}
D.~Carney, U.~\c{C}etintemel, M.~Cherniack, C.~Convey, S.~Lee, G.~Seidman,
  M.~Stonebraker, N.~Tatbul, and S.~Zdonik.
\newblock Monitoring streams: A new class of data management applications.
\newblock In {\em Proceedings of the 28th International Conference on Very
  Large Data Bases}, pages 215--226, Hong Kong, China, 2002.

\bibitem{Chabchoub10}
Y.~Chabchoub and G.~Hebrail.
\newblock Sliding {HyperLogLog}: Estimating cardinality in a data stream over a
  sliding window.
\newblock In {\em Proceedings of the IEEE International Conference on Data
  Mining Workshops}, pages 1297--1303, 2010.

\bibitem{Chandrasekaran03}
S.~Chandrasekaran, O.~Cooper, A.~Deshpande, M.~J. Franklin, J.~M. Hellerstein,
  W.~Hong, S.~Krishnamurthy, S.~R. Madden, F.~Reiss, and M.~A. Shah.
\newblock {TelegraphCQ}: Continuous dataflow processing.
\newblock In {\em Proceedings of the 2003 ACM SIGMOD International Conference
  on Management of Data}, pages 668--668, 2003.

\bibitem{Chang06}
F.~Chang, J.~Dean, S.~Ghemawat, W.~C. Hsieh, D.~A. Wallach, M.~Burrows,
  T.~Chandra, A.~Fikes, and R.~E. Gruber.
\newblock Bigtable: A distributed storage system for structured data.
\newblock In {\em Proceedings of the 7th Symposium on Operating Systems Design
  and Implementation}, pages 205--218, 2006.

\bibitem{Charikar04}
M.~Charikar, K.~Chen, and M.~Farach-Colton.
\newblock Finding frequent items in data streams.
\newblock {\em Theoretical Computer Science}, 312(1):3--15, 2004.

\bibitem{Chaudhuri99}
S.~Chaudhuri, R.~Motwani, and V.~Narasayya.
\newblock On random sampling over joins.
\newblock In {\em Proceedings of the 1999 ACM SIGMOD International Conference
  on Management of Data}, pages 263--274, Philadelphia, PA, 1999.

\bibitem{Chen11}
A.~Chen and J.~Cao.
\newblock Distinct counting with a self-learning bitmap.
\newblock In {\em Data Engineering, 2009. ICDE '09. IEEE 25th International
  Conference on}, pages 1171--1174, March 2009.

\bibitem{Chen07}
Y.~Chen, M.~Nascimento, B.~C. Ooi, and A.~Tung.
\newblock Spade: On shape-based pattern detection in streaming time series.
\newblock In {\em Proceedings of the IEEE 23rd International Conference on Data
  Engineering}, pages 786--795, April 2007.

\bibitem{Chitnis15}
R.~Chitnis, G.~Cormode, M.~Hajiaghayi, and M.~Monemizadeh.
\newblock Parameterized streaming: Maximal matching and vertex cover.
\newblock In {\em Proceedings of the Twenty-Sixth Annual ACM-SIAM Symposium on
  Discrete Algorithms}, pages 1234--1251, 2015.

\bibitem{Cohen03}
S.~Cohen and Y.~Matias.
\newblock Spectral bloom filters.
\newblock In {\em Proceedings of the 2003 ACM SIGMOD International Conference
  on Management of Data}, pages 241--252, 2003.

\bibitem{Coppersmith04}
D.~Coppersmith and R.~Kumar.
\newblock An improved data stream algorithm for frequency moments.
\newblock In {\em Proceedings of the Fifteenth Annual ACM-SIAM Symposium on
  Discrete Algorithms}, pages 151--156, New Orleans, Louisiana, 2004.

\bibitem{HealthCare1}
K.~Corbin.
\newblock {How CIOs Can Prepare for Healthcare `Data Tsunami’ More like
  this}.
\newblock
  \url{http://www.cio.com/article/2860072/healthcare/how-cios-can-prepare-for-healthcare-data-tsunami.html}.

\bibitem{Cormode08}
G.~Cormode and M.~Hadjieleftheriou.
\newblock Finding frequent items in data streams.
\newblock {\em Proceedings of the VLDB Endowment}, 1(2):1530--1541, Aug. 2008.

\bibitem{Cormode03}
G.~Cormode, F.~Korn, S.~Muthukrishnan, and D.~Srivastava.
\newblock Finding hierarchical heavy hitters in data streams.
\newblock In {\em Proceedings of the 29th International Conference on Very
  Large Data Bases - Volume 29}, pages 464--475, Berlin, Germany, 2003.

\bibitem{Cormode05}
G.~Cormode and S.~Muthukrishnan.
\newblock An improved data stream summary: The count-min sketch and its
  applications.
\newblock {\em Journal of Algorithms}, 55(1):58--75, Apr. 2005.

\bibitem{Cormode05a}
G.~Cormode, S.~Muthukrishnan, and I.~Rozenbaum.
\newblock Summarizing and mining inverse distributions on data streams via
  dynamic inverse sampling.
\newblock In {\em Proceedings of the 31st International Conference on Very
  Large Data Bases}, pages 25--36, Trondheim, Norway, 2005.

\bibitem{Cormode10}
G.~Cormode, S.~Muthukrishnan, K.~Yi, and Q.~Zhang.
\newblock Optimal sampling from distributed streams.
\newblock In {\em Proceedings of the Twenty-ninth ACM SIGMOD-SIGACT-SIGART
  Symposium on Principles of Database Systems}, pages 77--86, Indianapolis,
  Indiana, USA, 2010.

\bibitem{Cormode12}
G.~Cormode, S.~Muthukrishnan, K.~Yi, and Q.~Zhang.
\newblock Continuous sampling from distributed streams.
\newblock {\em Journal of the ACM}, 59(2):10:1--10:25, May 2012.

\bibitem{Dasu09}
T.~Dasu, S.~Krishnan, D.~Lin, S.~Venkatasubramanian, and K.~Yi.
\newblock Change (detection) you can believe in: Finding distributional shifts
  in data streams.
\newblock In {\em Proceedings of the 8th International Symposium on Intelligent
  Data Analysis: Advances in Intelligent Data Analysis VIII}, pages 21--34,
  2009.

\bibitem{Datar02}
M.~Datar and S.~Muthukrishnan.
\newblock Estimating rarity and similarity over data stream windows.
\newblock In {\em Proceedings of the 10th Annual European Symposium on
  Algorithms}, pages 323--334, 2002.

\bibitem{Dautrich13}
J.~L. Dautrich, Jr. and C.~V. Ravishankar.
\newblock Inferential time-decaying bloom filters.
\newblock In {\em Proceedings of the 16th International Conference on Extending
  Database Technology}, pages 239--250, Genoa, Italy, 2013.

\bibitem{Dean04}
J.~Dean and S.~Ghemawat.
\newblock Mapreduce: simplified data processing on large clusters.
\newblock In {\em Proceedings of the 6th conference on Symposium on Opearting
  Systems Design \& Implementation - Volume 6}, pages 10--10, San Francisco,
  CA, 2004.

\bibitem{Demaine02}
E.~D. Demaine, A.~L\'{o}pez-Ortiz, and J.~I. Munro.
\newblock Frequency estimation of internet packet streams with limited space.
\newblock In {\em Proceedings of the 10th Annual European Symposium on
  Algorithms}, pages 348--360, 2002.

\bibitem{Donnet06}
B.~Donnet, B.~Baynat, and T.~Friedman.
\newblock Retouched bloom filters: Allowing networked applications to trade off
  selected false positives against false negatives.
\newblock In {\em Proceedings of the 2006 ACM CoNEXT Conference}, pages
  13:1--13:12, Lisboa, Portugal, 2006.

\bibitem{Teixeira10}
P.~H. dos Santos~Teixeira and R.~L. Milidi\'{u}.
\newblock Data stream anomaly detection through principal subspace tracking.
\newblock In {\em Proceedings of the 2010 ACM Symposium on Applied Computing},
  pages 1609--1616, Sierre, Switzerland, 2010.

\bibitem{Durand03}
M.~Durand and P.~Flajolet.
\newblock {LogLog Counting of Large Cardinalities}.
\newblock In {\em Annual European Symposium on Algorithms, volume 2832 of
  LNCS}, pages 605--617, 2003.

\bibitem{Eppstein97}
D.~Eppstein, Z.~Galil, G.~F. Italiano, and A.~Nissenzweig.
\newblock Sparsification -- a technique for speeding up dynamic graph
  algorithms.
\newblock {\em Journal of the ACM}, 44(5):669--696, Sept. 1997.

\bibitem{Esfandiari15}
H.~Esfandiari, M.~T. Hajiaghayi, V.~Liaghat, M.~Monemizadeh, and K.~Onak.
\newblock Streaming algorithms for estimating the matching size in planar
  graphs and beyond.
\newblock In {\em Proceedings of the Twenty-Sixth Annual ACM-SIAM Symposium on
  Discrete Algorithms}, pages 1217--1233, 2015.

\bibitem{Estan02}
C.~Estan and G.~Varghese.
\newblock New directions in traffic measurement and accounting.
\newblock In {\em Proceedings of the 2002 Conference on Applications,
  Technologies, Architectures, and Protocols for Computer Communications},
  pages 323--336, Pittsburgh, Pennsylvania, USA, 2002.

\bibitem{Fan14}
B.~Fan, D.~G. Andersen, M.~Kaminsky, and M.~D. Mitzenmacher.
\newblock Cuckoo filter: Practically better than bloom.
\newblock In {\em Proceedings of the 10th ACM International on Conference on
  Emerging Networking Experiments and Technologies}, pages 75--88, Sydney,
  Australia, 2014.

\bibitem{Feigenbaum05}
J.~Feigenbaum, S.~Kannan, A.~McGregor, S.~Suri, and J.~Zhang.
\newblock On graph problems in a semi-streaming model.
\newblock {\em Journal of Theoretical Computer Science}, 348(2):207--216, Dec.
  2005.

\bibitem{Feigenblat10}
G.~Feigenblat, O.~Itzhaki, and E.~Porat.
\newblock The frequent items problem, under polynomial decay, in the streaming
  model.
\newblock {\em Theoretical Computer Science}, 411(34-36):3048--3054, July 2010.

\bibitem{Flajolet07}
P.~Flajolet, E.~Fusy, O.~Gandouet, and F.~Meunier.
\newblock Hyperloglog: The analysis of a near-optimal cardinality estimation
  algorithm.
\newblock In {\em PROCEEDINGS OF THE 2007 INTERNATIONAL CONFERENCE ON ANALYSIS
  OF ALGORITHMS}, 2007.

\bibitem{Flajolet83}
P.~Flajolet and G.~N. Martin.
\newblock Probabilistic counting.
\newblock In {\em Proceedings of the 24th Annual Symposium on Foundations of
  Computer Science}, pages 76--82, Washington, DC, USA, 1983.

\bibitem{Gal10}
A.~G\'{a}l and P.~Gopalan.
\newblock Lower bounds on streaming algorithms for approximating the length of
  the longest increasing subsequence.
\newblock {\em SIAM Journal on Computing}, 39(8):3463--3479, Aug. 2010.

\bibitem{Gandhi07}
S.~Gandhi, S.~Suri, and E.~Welzl.
\newblock Catching elephants with mice: Sparse sampling for monitoring sensor
  networks.
\newblock In {\em Proceedings of the 5th International Conference on Embedded
  Networked Sensor Systems}, pages 261--274, Sydney, Australia, 2007.

\bibitem{Ganguly07}
S.~Ganguly.
\newblock Counting distinct items over update streams.
\newblock {\em Theoretical Computer Science}, 378(3):211--222, June 2007.

\bibitem{Gemulla08}
R.~Gemulla and W.~Lehner.
\newblock Sampling time-based sliding windows in bounded space.
\newblock In {\em Proceedings of the 2008 ACM SIGMOD International Conference
  on Management of Data}, pages 379--392, Vancouver, Canada, 2008.

\bibitem{Gilbert02}
A.~C. Gilbert, S.~Guha, P.~Indyk, Y.~Kotidis, S.~Muthukrishnan, and M.~J.
  Strauss.
\newblock Fast, small-space algorithms for approximate histogram maintenance.
\newblock In {\em Proceedings of the ACM Symposium on Theory of Computing},
  pages 389--398, Montreal, Quebec, Canada, 2002.

\bibitem{Giroire05}
F.~Giroire.
\newblock Order statistics and estimating cardinalities of massive data sets.
\newblock In {\em Proceedings of the 2005 International Conference on Analysis
  of Algorithms}, pages 157--166, 2005.

\bibitem{Greenwald01}
M.~Greenwald and S.~Khanna.
\newblock Space-efficient online computation of quantile summaries.
\newblock In {\em Proceedings of the 2001 ACM SIGMOD International Conference
  on Management of Data}, pages 58--66, Santa Barbara, California, USA, 2001.

\bibitem{Gualtieri14}
M.~Gualtieri and R.~Curran.
\newblock {The Forrester Wave\texttrademark: Big Data Streaming Analytics
  Platforms, Q3 2014}.
\newblock 2014.

\bibitem{Guha09a}
S.~Guha.
\newblock Tight results for clustering and summarizing data streams.
\newblock In {\em Proceedings of the 12th International Conference on Database
  Theory}, pages 268--275, St. Petersburg, Russia, 2009.

\bibitem{Guha06}
S.~Guha, N.~Koudas, and K.~Shim.
\newblock Approximation and streaming algorithms for histogram construction
  problems.
\newblock {\em ACM Transactions on Database Systems}, 31(1):396--438, Mar.
  2006.

\bibitem{Guha09}
S.~Guha and A.~McGregor.
\newblock Stream order and order statistics: Quantile estimation in
  random-order streams.
\newblock {\em SIAM Journal on Computing}, 38(5):2044--2059, 2009.

\bibitem{Guha00}
S.~Guha, N.~Mishra, R.~Motwani, and L.~O'Callaghan.
\newblock Clustering data streams.
\newblock In {\em Proceedings of the 41st Annual Symposium on Foundations of
  Computer Science}, pages 359--366, 2000.

\bibitem{Guo14}
T.~Guo, S.~Sathe, and K.~Aberer.
\newblock Fast correlation discovery for large-scale streaming time-series
  data.
\newblock Technical Report EPFL-REPORT-200473, EPFL, 2014.

\bibitem{Halatchev05}
M.~Halatchev and L.~Gruenwald.
\newblock Estimating missing values in related sensor data streams.
\newblock In {\em Proceedings of the 11th International Conference on
  Management of Data}, pages 83--94, Hyderabad, India, 2005.

\bibitem{Halldorsson10}
B.~V. Halld\'{o}rsson, M.~M. Halld\'{o}rsson, E.~Losievskaja, and M.~Szegedy.
\newblock Streaming algorithms for independent sets.
\newblock In {\em Proceedings of the 37th International Colloquium Conference
  on Automata, Languages and Programming}, pages 641--652, Bordeaux, France,
  2010.

\bibitem{Hao07}
F.~Hao, M.~Kodialam, and T.~V. Lakshman.
\newblock Building high accuracy bloom filters using partitioned hashing.
\newblock In {\em Proceedings of the 2007 ACM SIGMETRICS International
  Conference on Measurement and Modeling of Computer Systems}, pages 277--288,
  2007.

\bibitem{Heule13}
S.~Heule, M.~Nunkesser, and A.~Hall.
\newblock Hyperloglog in practice: Algorithmic engineering of a state of the
  art cardinality estimation algorithm.
\newblock In {\em Proceedings of the 16th International Conference on Extending
  Database Technology}, pages 683--692, Genoa, Italy, 2013.

\bibitem{Homem10}
N.~Homem and J.~P. Carvalho.
\newblock Finding top-k elements in data streams.
\newblock {\em Information Sciences}, 180(24):4958 -- 4974, 2010.

\bibitem{Hore07}
P.~Hore, L.~Hall, and D.~Goldgof.
\newblock Creating streaming iterative soft clustering algorithms.
\newblock In {\em Annual Meeting of the North American Fuzzy Information
  Processing Society}, pages 484--488, June 2007.

\bibitem{Hung10a}
R.~Y.~S. Hung, L.-K. Lee, and H.~F. Ting.
\newblock Finding frequent items over sliding windows with constant update
  time.
\newblock {\em Information Processing Letters}, 110(7):257--260, Mar. 2010.

\bibitem{Hung10}
R.~Y.~S. Hung and H.~F. Ting.
\newblock An $\omega(1/\epsilon \log 1/\epsilon)$ space lower bound for finding
  $\epsilon$-approximate quantiles in a data stream.
\newblock In {\em Proceedings of the 4th International Conference on Frontiers
  in Algorithmics}, pages 89--100, Wuhan, China, 2010.

\bibitem{Hwang05}
J.-H. Hwang, M.~Balazinska, A.~Rasin, U.~Cetintemel, M.~Stonebraker, and
  S.~Zdonik.
\newblock High-availability algorithms for distributed stream processing.
\newblock In {\em Proceedings of the International Conference on Data
  Engineering}, pages 779--790, April 2005.

\bibitem{Indyk05}
P.~Indyk and D.~Woodruff.
\newblock Optimal approximations of the frequency moments of data streams.
\newblock In {\em Proceedings of the Thirty-seventh Annual ACM Symposium on
  Theory of Computing}, pages 202--208, Baltimore, MD, 2005.

\bibitem{Jin03}
C.~Jin, W.~Qian, C.~Sha, J.~X. Yu, and A.~Zhou.
\newblock Dynamically maintaining frequent items over a data stream.
\newblock In {\em Proceedings of the Twelfth International Conference on
  Information and Knowledge Management}, pages 287--294, New Orleans, LA, 2003.

\bibitem{Kalman60}
R.~E. Kalman.
\newblock A new approach to linear filtering and prediction problems.
\newblock {\em Journal of Basic Engineering}, 82(Series D):35--45, 1960.

\bibitem{Kane10}
D.~M. Kane, J.~Nelson, and D.~P. Woodruff.
\newblock An optimal algorithm for the distinct elements problem.
\newblock In {\em Proceedings of the Twenty-ninth ACM SIGMOD-SIGACT-SIGART
  Symposium on Principles of Database Systems}, pages 41--52, Indianapolis,
  Indiana, USA, 2010.

\bibitem{Kapralov14}
M.~Kapralov, S.~Khanna, and M.~Sudan.
\newblock Approximating matching size from random streams.
\newblock In {\em Proceedings of the Twenty-Fifth Annual ACM-SIAM Symposium on
  Discrete Algorithms}, pages 734--751, Portland, OR, 2014.

\bibitem{Karp03}
R.~M. Karp, S.~Shenker, and C.~H. Papadimitriou.
\newblock A simple algorithm for finding frequent elements in streams and bags.
\newblock {\em ACM Transactions on Database Systems}, 28(1):51--55, Mar. 2003.

\bibitem{Khan09}
M.~Khan.
\newblock Anomaly detection in data streams using fuzzy logic.
\newblock In {\em Proceedings of the International Conference on Information
  and Communication Technologies}, pages 167--174, Aug 2009.

\bibitem{Kirsch08}
A.~Kirsch and M.~Mitzenmacher.
\newblock Less hashing, same performance: Building a better bloom filter.
\newblock {\em Random Struct. Algorithms}, 33(2):187--218, Sept. 2008.

\bibitem{Koper11}
A.~Koper and H.~Nguyen.
\newblock Sequential pattern mining from stream data.
\newblock In J.~Tang, I.~King, L.~Chen, and J.~Wang, editors, {\em Advanced
  Data Mining and Applications}, volume 7121 of {\em Lecture Notes in Computer
  Science}, pages 278--291. 2011.

\bibitem{Kulkarni15}
S.~Kulkarni, N.~Bhagat, M.~Fu, V.~Kedigehalli, C.~Kellogg, S.~Mittal, J.~M.
  Patel, K.~Ramasamy, and S.~Taneja.
\newblock Twitter {H}eron: Streaming at scale.
\newblock In {\em Proceedings of SIGMOD}, Melbourne, Australia, 2015.

\bibitem{Lee06}
L.~K. Lee and H.~F. Ting.
\newblock Maintaining significant stream statistics over sliding windows.
\newblock In {\em Proceedings of the Seventeenth Annual ACM-SIAM Symposium on
  Discrete Algorithm}, pages 724--732, Miami, FL, 2006.

\bibitem{Li04}
H.-F. Li, S.-Y. Lee, and M.-K. Shan.
\newblock Mining frequent closed structures in streaming melody sequences.
\newblock In {\em IEEE International Conference on Multimedia and Expo},
  volume~3, pages 2031--2034, June 2004.

\bibitem{Li09}
L.~Li, J.~McCann, N.~S. Pollard, and C.~Faloutsos.
\newblock {DynaMMo}: Mining and summarization of coevolving sequences with
  missing values.
\newblock In {\em Proceedings of the 15th ACM SIGKDD International Conference
  on Knowledge Discovery and Data Mining}, KDD '09, pages 507--516, Paris,
  France, 2009.

\bibitem{Nowell05}
D.~Liben-Nowell, E.~Vee, and A.~Zhu.
\newblock Finding longest increasing and common subsequences in streaming data.
\newblock In {\em Proceedings of the 11th Annual International Conference on
  Computing and Combinatorics}, pages 263--272, Kunming, China, 2005.

\bibitem{Ma13}
Q.~Ma, S.~Muthukrishnan, and M.~Sandler.
\newblock Frugal streaming for estimating quantiles.
\newblock In A.~Brodnik, A.~López-Ortiz, V.~Raman, and A.~Viola, editors, {\em
  Space-Efficient Data Structures, Streams, and Algorithms}, volume 8066 of
  {\em Lecture Notes in Computer Science}, pages 77--96. 2013.

\bibitem{Manerikar09}
N.~Manerikar and T.~Palpanas.
\newblock Frequent items in streaming data: An experimental evaluation of the
  state-of-the-art.
\newblock {\em Data \& Knowledge Engineering}, 68(4):415--430, Apr. 2009.

\bibitem{Manku02}
G.~S. Manku and R.~Motwani.
\newblock Approximate frequency counts over data streams.
\newblock In {\em Proceedings of the 28th International Conference on Very
  Large Data Bases}, pages 346--357, Hong Kong, China, 2002.

\bibitem{Manyika11}
J.~Manyika, M.~Chui, B.~Brown, J.~Bughin, R.~Dobbs, C.~Roxburgh, and A.~H.
  Byers.
\newblock Big data: The next frontier for innovation, competition, and
  productivity.
\newblock
  \url{http://www.mckinsey.com/Insights/MGI/Research/Technology_and_Innovation/Big_data_The_next_frontier_for_innovation},
  May 2011.

\bibitem{McGregor14}
A.~McGregor.
\newblock Graph stream algorithms: A survey.
\newblock {\em SIGMOD Record}, 43(1):9--20, May 2014.

\bibitem{Metwally05}
A.~Metwally, D.~Agrawal, and A.~El~Abbadi.
\newblock Efficient computation of frequent and top-k elements in data streams.
\newblock In T.~Eiter and L.~Libkin, editors, {\em Database Theory - ICDT
  2005}, volume 3363 of {\em Lecture Notes in Computer Science}, pages
  398--412. 2005.

\bibitem{Moraru12}
I.~Moraru and D.~G. Andersen.
\newblock Exact pattern matching with feed-forward bloom filters.
\newblock {\em Journal on Experimental Algorithmics}, 17:3.4:3.1--3.4:3.18,
  Sept. 2012.

\bibitem{Pulsar}
S.~Murthy and T.~Ng.
\newblock {Announcing Pulsar: Real-time Analytics at Scale}.
\newblock
  \url{http://www.ebaytechblog.com/2015/02/23/announcing-pulsar-real-time-analytics-at-scale},
  Feb. 2015.

\bibitem{S4}
L.~Neumeyer, B.~Robbins, A.~Nair, and A.~Kesari.
\newblock {S4}: Distributed stream computing platform.
\newblock In {\em Proceedings of the 2010 IEEE International Conference on Data
  Mining Workshops}, pages 170--177, 2010.
\newblock \url{http://incubator.apache.org/s4/}.

\bibitem{Callaghan02}
L.~O'Callaghan, N.~Mishra, A.~Meyerson, S.~Guha, and R.~Motwani.
\newblock Streaming-data algorithms for high-quality clustering.
\newblock In {\em Proceedings of the 18th International Conference on Data
  Engineering}, pages 685--694, 2002.

\bibitem{Pagh05}
A.~Pagh, R.~Pagh, and S.~S. Rao.
\newblock An optimal bloom filter replacement.
\newblock In {\em Proceedings of the Sixteenth Annual ACM-SIAM Symposium on
  Discrete Algorithms}, pages 823--829, Vancouver, BC, 2005.

\bibitem{Pan12}
S.~Pan and X.~Zhu.
\newblock Cgstream: Continuous correlated graph query for data streams.
\newblock In {\em Proceedings of the 21st ACM International Conference on
  Information and Knowledge Management}, pages 1183--1192, Maui, Hawaii, USA,
  2012.

\bibitem{Papadimitriou05}
S.~Papadimitriou, J.~Sun, and C.~Faloutsos.
\newblock Streaming pattern discovery in multiple time-series.
\newblock In {\em Proceedings of the 31st International Conference on Very
  Large Data Bases}, pages 697--708, Trondheim, Norway, 2005.

\bibitem{Pentland13}
A.~S. Pentland.
\newblock The data-driven society.
\newblock {\em Scientific American}, 309:78--83, 2013.

\bibitem{Pripuzic15}
K.~Pripu\v{z}i\'{c}, I.~P. \v{Z}arko, and K.~Aberer.
\newblock Time- and space-efficient sliding window top-k query processing.
\newblock {\em ACM Trans. Database Syst.}, 40(1):1:1--1:44, Mar. 2015.

\bibitem{Putze10}
F.~Putze, P.~Sanders, and J.~Singler.
\newblock Cache-, hash-, and space-efficient bloom filters.
\newblock {\em Journal of Experimental Algorithmics}, 14:4:4.4--4:4.18, Jan.
  2010.

\bibitem{Raissi08}
C.~Ra\"{i}ssi and M.~Plantevit.
\newblock Mining multidimensional sequential patterns over data streams.
\newblock In I.-Y. Song, J.~Eder, and T.~Nguyen, editors, {\em Data Warehousing
  and Knowledge Discovery}, volume 5182 of {\em Lecture Notes in Computer
  Science}, pages 263--272. 2008.

\bibitem{Rashidi10}
P.~Rashidi and D.~Cook.
\newblock Mining sensor streams for discovering human activity patterns over
  time.
\newblock In {\em Proceedings of the IEEE 10th International Conference on Data
  Mining}, pages 431--440, Dec 2010.

\bibitem{Rhodes15}
L.~Rhodes.
\newblock {DataSketches: Fast, Approximate Analysis of Big Data}.
\newblock \url{https://yahooeng.tumblr.com/post/135390948446/data-sketches},
  Dec. 2015.

\bibitem{Rodrigues06}
P.~P. Rodrigues and J.~Gama.
\newblock Online prediction of streaming sensor data.
\newblock In {\em Proceedings of the 3rd International Wokshop on Knowledge
  Discovery from Data Streams}, 2006.

\bibitem{Rothenberg10}
C.~E. Rothenberg, C.~A.~B. Macapuna, F.~L. Verdi, and M.~F. Magalh\~{a}es.
\newblock The deletable bloom filter: A new member of the bloom family.
\newblock {\em IEEE Communications Letters}, 14(6):557--559, June 2010.

\bibitem{Rottenstreich14}
O.~Rottenstreich, Y.~Kanizo, and I.~Keslassy.
\newblock The variable-increment counting bloom filter.
\newblock {\em IEEE/ACM Transactions on Networking}, 22(4):1092--1105, Aug.
  2014.

\bibitem{Sahpaski12}
D.~Sahpaski, A.~S. Dimovski, G.~Velinov, and M.~Kon-Popovska.
\newblock Efficient processing of {Top-k J}oin queries by attribute domain
  refinement.
\newblock In {\em Proceedings of the 16th East European Conference on Advances
  in Databases and Information Systems}, pages 318--331, Pozna\&\#324;, Poland,
  2012.

\bibitem{Sayal04}
M.~Sayal.
\newblock Detecting time correlations in time--series data streams.
\newblock Technical Report HPL-2004-103, 2004.

\bibitem{HealthCare}
N.~D. Shah and J.~Pathak.
\newblock {Why Health Care May Finally Be Ready for Big Data}.
\newblock
  \url{https://hbr.org/2014/12/why-health-care-may-finally-be-ready-for-big-data}.

\bibitem{Shrivastava04}
N.~Shrivastava, C.~Buragohain, D.~Agrawal, and S.~Suri.
\newblock Medians and beyond: New aggregation techniques for sensor networks.
\newblock In {\em Proceedings of the 2Nd International Conference on Embedded
  Networked Sensor Systems}, pages 239--249, Baltimore, MD, USA, 2004.

\bibitem{Silva13}
J.~A. Silva, E.~R. Faria, R.~C. Barros, E.~R. Hruschka, A.~C. P. L. F.~d.
  Carvalho, and J.~a. Gama.
\newblock Data stream clustering: A survey.
\newblock {\em ACM Computing Surveys}, 46(1):13:1--13:31, July 2013.

\bibitem{Su07}
L.~Su, W.~Han, S.~Yang, P.~Zou, and Y.~Jia.
\newblock Continuous adaptive outlier detection on distributed data streams.
\newblock In {\em Proceedings of the Third International Conference on High
  Performance Computing and Communications}, pages 74--85, 2007.

\bibitem{Subramaniam06}
S.~Subramaniam, T.~Palpanas, D.~Papadopoulos, V.~Kalogeraki, and D.~Gunopulos.
\newblock Online outlier detection in sensor data using non-parametric models.
\newblock In {\em Proceedings of the 32Nd International Conference on Very
  Large Data Bases}, pages 187--198, Seoul, Korea, 2006.

\bibitem{Sun07}
X.~Sun and D.~P. Woodruff.
\newblock The communication and streaming complexity of computing the longest
  common and increasing subsequences.
\newblock In {\em Proceedings of the Eighteenth Annual ACM-SIAM Symposium on
  Discrete Algorithms}, SODA '07, pages 336--345, New Orleans, Louisiana, 2007.

\bibitem{Tan11}
S.~C. Tan, K.~M. Ting, and T.~F. Liu.
\newblock Fast anomaly detection for streaming data.
\newblock In {\em Proceedings of the Twenty-Second International Joint
  Conference on Artificial Intelligence - Volume Volume Two}, IJCAI'11, pages
  1511--1516, Barcelona, Spain, 2011.

\bibitem{Tanbeer09}
S.~K. Tanbeer, C.~F. Ahmed, B.-S. Jeong, and Y.-K. Lee.
\newblock Efficient single-pass frequent pattern mining using a prefix-tree.
\newblock {\em Information Sciences}, 179(5):559 -- 583, 2009.

\bibitem{Tanbeer09a}
S.~K. Tanbeer, C.~F. Ahmed, B.-S. Jeong, and Y.-K. Lee.
\newblock Sliding window-based frequent pattern mining over data streams.
\newblock {\em Information Sciences}, 179(22):3843 -- 3865, 2009.

\bibitem{Tao07}
Y.~Tao, X.~Lian, D.~Papadias, and M.~Hadjieleftheriou.
\newblock Random sampling for continuous streams with arbitrary updates.
\newblock {\em IEEE Transactions on Knowledge and Data Engineering},
  19(1):96--110, Jan. 2007.

\bibitem{Ting14}
D.~Ting.
\newblock Streamed approximate counting of distinct elements: Beating optimal
  batch methods.
\newblock In {\em Proceedings of the 20th ACM SIGKDD International Conference
  on Knowledge Discovery and Data Mining}, pages 442--451, 2014.

\bibitem{Storm}
A.~Toshniwal, S.~Taneja, A.~Shukla, K.~Ramasamy, J.~M. Patel, S.~Kulkarni,
  J.~Jackson, K.~Gade, M.~Fu, J.~Donham, N.~Bhagat, S.~Mittal, and D.~Ryaboy.
\newblock Storm \ @ twitter.
\newblock In {\em Proceedings of the 2014 ACM SIGMOD International Conference
  on Management of Data}, pages 147--156, Snowbird, UT, 2014.
\newblock \url{https://storm.apache.org/}.

\bibitem{Toyoda13}
M.~Toyoda, Y.~Sakurai, and Y.~Ishikawa.
\newblock Pattern discovery in data streams under the time warping distance.
\newblock {\em The VLDB Journal}, 22(3):295--318, June 2013.

\bibitem{Vijayakumar07}
N.~N. Vijayakumar and B.~Plale.
\newblock Prediction of missing events in sensor data streams using kalman
  filters.
\newblock In {\em Proceedings of the First International Workshop on Knowledge
  Discovery from Sensor Data}, Aug. 2007.

\bibitem{Vitter85}
J.~S. Vitter.
\newblock Random sampling with a reservoir.
\newblock {\em ACM Transactions on Mathematical Software}, 11(1):37--57, Mar.
  1985.

\bibitem{Wan00}
E.~Wan and R.~Van Der~Merwe.
\newblock The unscented kalman filter for nonlinear estimation.
\newblock In {\em Proceedings of the Adaptive Systems for Signal Processing,
  Communications, and Control Symposium}, pages 153--158, 2000.

\bibitem{Wang03}
M.~Wang and X.~Wang.
\newblock Efficient evaluation of composite correlations for streaming time
  series.
\newblock In G.~Dong, C.~Tang, and W.~Wang, editors, {\em Advances in Web-Age
  Information Management}, pages 369--380. 2003.

\bibitem{Wang05}
Y.-l. Wang, H.-b. Xu, Y.-s. Dong, X.-j. Liu, and J.-b. Qian.
\newblock Apforecast: An adaptive forecasting method for data streams.
\newblock In R.~Khosla, R.~Howlett, and L.~Jain, editors, {\em Knowledge-Based
  Intelligent Information and Engineering Systems}, volume 3682 of {\em Lecture
  Notes in Computer Science}, pages 957--963. 2005.

\bibitem{Xie13}
Q.~Xie, S.~Shang, B.~Yuan, C.~Pang, and X.~Zhang.
\newblock Local correlation detection with linearity enhancement in streaming
  data.
\newblock In {\em Proceedings of the 22Nd ACM International Conference on
  Conference on Information \&\#38; Knowledge Management}, pages 309--318,
  2013.

\bibitem{Yang11}
D.~Yang, A.~Shastri, E.~A. Rundensteiner, and M.~O. Ward.
\newblock An optimal strategy for monitoring top-k queries in streaming
  windows.
\newblock In {\em Proceedings of the 14th International Conference on Extending
  Database Technology}, pages 57--68, Uppsala, Sweden, 2011.

\bibitem{Yang12}
F.~Yang, Z.~Qian, X.~Chen, I.~Beschastnikh, L.~Zhuang, L.~Zhou, and G.~Shen.
\newblock Sonora: A platform for continuous mobile-cloud computing.
\newblock Technical Report MSR-TR-2012-34, Microsoft Research Asia, 2012.

\bibitem{Yu12}
K.~Yu, W.~Ding, D.~A. Simovici, and X.~Wu.
\newblock Mining emerging patterns by streaming feature selection.
\newblock In {\em Proceedings of the 18th ACM SIGKDD International Conference
  on Knowledge Discovery and Data Mining}, pages 60--68, Beijing, China, 2012.

\bibitem{Zhang05}
L.~Zhang, Z.~Li, M.~Yu, Y.~Wang, and Y.~Jiang.
\newblock Random sampling algorithms for sliding windows over data streams.
\newblock In {\em Proceedings of the 11th Joint International Computer
  Conference}, pages 572--575, 2005.

\bibitem{Zhang07}
Q.~Zhang and W.~Wang.
\newblock An efficient algorithm for approximate biased quantile computation in
  data streams.
\newblock In {\em Proceedings of the Sixteenth ACM Conference on Conference on
  Information and Knowledge Management}, pages 1023--1026, Lisbon, Portugal,
  2007.

\bibitem{Zhang13}
Y.~Zhang and L.~Liu.
\newblock Distance-aware bloom filters: Enabling collaborative search for
  efficient resource discovery.
\newblock {\em Journal on Future Generation Computer Systems},
  29(6):1621--1630, Aug. 2013.

\end{thebibliography}

}

\end{document}